\documentclass[a4paper,12pt]{article}
\usepackage{amssymb,amsfonts,amsmath,mathtext,cite,enumerate,float}
\usepackage{amssymb,amsthm,latexsym,euscript,amscd, authblk}
\usepackage{braket}

\usepackage[english]{babel}
\usepackage[cp1251]{inputenc}

\title{On non-commutative operator graphs generated by reducible unitary representation of the Heisenberg-Weyl group}

\author[1,2]{G.G.Amosov\thanks {gramos@mi-ras.ru}}

\author[2]{A.S. Mokeev\thanks {alexandrmokeev@yandex.ru}}

\affil[1]{Steklov Mathematical Institute of Russian Academy of Sciences, 8 Gubkina St., Moscow, 119991 Russia}

\affil[2]{St. Petersburg State University, 7/9 Universitetskaya nab., St. Petersburg, 199034
Russia}

\begin{document}

\maketitle

\begin{abstract}
We consider a reducible unitary representation of Heisenberg-Weyl group in a tensor product of two Hilbert spaces. 
A non-commutative operator graph generated by this representation is introduced. It is shown that spectral projections of unitaries in the representation are anticliques (quantum error-correcting codes) for this graph. The obtained codes are appeared to be linear envelopes of entangled vectors.
\end{abstract}


\section{Introduction}

The theory of non-commutative operator graphs \cite{Winter} plays a central role in the study of the conditions under which it is possible to encode information such that the transmission is made with zero error. Herewith such tasks as estimating the maximal dimension of graph \cite{amosovmokeev} as well as constructing examples of classical \cite {cubitt} and quantum  \cite {shirokov, shirokov2} superactivation appear.
Recently, a study of noncommutative operator graphs generated by covariant resolutions of identity was initiated \cite{amo}. In \cite{amosovmokeev2} a case of the circle group which is compact and commutative was investigated in detail. In this paper we introduce graphs generated by a reducible unitary representation of the discrete Heisenberg-Weyl group in a tensor product of two Hilbert spaces.
It gives a more deep example with respect to \cite{amosovmokeev2} because the Heisenber-Weyl group is non-commutative.

As is known, there is a direct connection between the theory of quantum error-correcting codes and the theory of non-commutative operator graphs (see brief introduction in \cite{amosovmokeev} or a more thorough statement of the problem in \cite{hol2,laff}). The theory of covariant resolutions of identity is detailed in \cite{amosovmokeev2}.

Non-commutative operator graph is a subspace $\mathcal{V}$ in the algebra of all bounded linear operators $B(H)$ in a Hilbert space $H$ closed under operator conjugation and containing the identity operator. A graph $\mathcal{V}$ is said to be satisfying the Knill-Laflamme condition \cite{laff} if there is a orthogonal projection $P_K,\ rankP_L\ge 2,$ on the subspace $K\subset H$ such that
$$
P_K \mathcal{V} P_K=\mathbb{C} P_K.
$$
The projection $P_K$ and the space $K$ are said to be an anticlique \cite {Weaver} and a quantum error-correcting code correspondingly.
Consider a locally compact group $G$ with the Haar measure $\mu$ normalised by the condition $\mu(G)=1$. Let $\mathcal{B}$ be a $\sigma$-algebra generated by all compact subsets of $G$. The set $\lbrace M(B), B \in \mathcal{B} \rbrace$ containing positive operators in $H$ is said to be a resolution of identity if \cite{hol1}
$$
M(\emptyset)=0, M(G)=I,
$$

$$
M(\cup_j B_j)=\sum_j M(B_j), \ B_k\cap B_m=\emptyset \  \textit{for}\  k\neq m; B_k \in \mathcal{B}.
$$   

For a projective unitary representation $\rho:g \mapsto  U_g$ of the group $G$ we define an action of $g \in G$ on operator $S \in B(H)$ by the rule
$$
\rho_g(S)=U_g S U_g^{*}.
$$ 
A resolution of identity is said to be covariant with respect to this action if
$$
\rho_g(M(B))=M(gB), \forall g \in G, \ \forall B \in \mathcal{B}.
$$
In a finite dimensional case each covariant resolution of identity is known to have the form
$$
M(B)=\int_B U_g M_0 U_g^{*} d \mu(g)
$$
where $M_0$ is some positive operator.

Consider the graph $\mathcal{V}$ generated by a covariant resolution of identity $\lbrace M(B), B \in \mathcal{B} \rbrace$, i. e.
$$
\mathcal{V}= span \lbrace M(B), B \in \mathcal{B} \rbrace.
$$
For such graphs it is possible to find sufficient conditions for the existence of anticliques. Consider the spectral decomposition of a unitary operators $U_g$ 
$$
U_g=\sum_{j\in J_g} a_{j} (g) P^{g}_{j}.
$$

\textbf{Proposition 1 \cite{amosovmokeev2}.}\textit{ Suppose that given $g\in G$ there exists $j_g\in J_g$ such that $P_{j_g}^g=P$ and $rankP\ge 2$. Then, $P$ is an anticlique for $\mathcal{V}$.}

Our previous examples of anticliques \cite {amosovmokeev2} were obtained due to application of Proposition 1 above.
In the present paper we don't build an example for which Proposition 1 gives an anticlique. Nevertheless the technique will be similar. The paper is organized as follows. In Section 2 we construct a reducible unitary representations of the Heisenberg-Weyl groups in a tensor product of two Hilbert spaces. Then, in Section 3 we obtain resolutions of identity covariant with respect to the action of representations constructed in Section 2. For the graphs generated by this resolutions of identity we prove that anticliques exist. In Section 4 a structure of this graphs is revealed.

\section{Reducible unitary representations of the Heisenberg - Weyl group}

Consider a Hilbert space $\mathfrak {H},\ dim\mathfrak {H}=n$, with the fixed orthogonal basis $\ket {j},\ 0\le j\le n-1$. Define two unitary operators $S$ and $M$
by the formula
$$
S\ket {j}=\ket {j+1\ mod\ n},\ M\ket {j}=e^{\frac {2\pi {\bf i}}{n}j}\ket {j}.
$$
The group $G_n$ generated by unitary operators $S$, $M$ and unimodular roots $\{e^{\frac {2\pi {\bf i}}{n}j},\ 0\le j\le n-1\}$ is known as the Heisenberg-Weyl group.
Below we shall construct the reducible unitary representation of $G_n$ in the Hilbert space $H=\mathfrak {H}\otimes \mathfrak {H}$.

Let $\ket{kj}, 0 \le k,j \le n-1,$ be the standard basis in $H$. 
Taking the discrete Fourier transform of $\ket {jj},\ 0\le j\le n-1,$ we get $n$ orthogonal vectors
\begin{equation}\label{bas1}
h_k^0=\frac{1}{\sqrt{n}}\sum_{j=0}^{n-1}e^{\frac {2\pi {\bf i}kj}{n}} \ket{jj},\ 0\le k\le n-1,
\end{equation}
Then, applying the shift operator we obtain other $n-1$ series of orthogonal vectors
\begin{equation}\label{bas2}
h_k^j=(I\otimes S^{j} )h_k^{0},\ 0\le k\le n-1,\ 1\le j\le n-1.
\end{equation}
Together $\{h_k^j,\ 0\le k,j\le n-1\}$ form the orthogonal basis in $H$ consisting of entangled vectors.
Consider the subspaces spanned by systems of the following entangled vectors
$$
H_j = span \{ h_k^j, 0 \le k \le n-1  \}.
$$
Let us define a map $\pi $ transmitting $S$ and $U$ to unitary operators in $H$
as follows
$$
\pi (S)h_k^j=h_{k+1\ mod\ n}^{j},
$$
$$
\pi (M)h_k^j=e^{\frac{2\pi {\bf i} }{n}k}h_k^{j}.
$$

{\bf Proposition.} {\it The map $\pi $ can be extended to a unitary representation of the group $G_n$.}

Proof.

Note that the subspaces $H_j$ are invariant with respect to the action of $S$ and $M$.
Let us extend $\pi $ to $G_n$ by the formula
$$
\pi (S^pM^q)=\pi (S)^p\pi (M)^q,
$$
then the restrictions of $\pi (G_n)$ to $H_j$ are unitary equivalent to $G_n$.

$\Box $

\section {Covariant resolutions of identity for $G_n$.}

Let us consider the conditional expectation (projection) on the algebra of stationary points with respect to the action
of the group $G_n$ determined by the representation $\pi$
\begin{equation}\label{proj}
{\mathbb E}(x)=\frac {1}{n^2}\sum \limits _{p,q=0}^{n-1}\pi (S)^p\pi (M)^qx\pi (M^q)^*\pi (S^p)^*,\ x\in B(H).
\end{equation}
Since $\pi $ is a direct sum of irreducible representations
$$
\pi =\oplus _j\pi |_{H_j},
$$
the algebra of elements which are stationary with respect to the action of $G_n$ has the generators
$$
x_{pq}=\sum \limits _{k=0}^{n-1}\ket {h_k^p}\bra {h_k^q},
$$
$0\le p,q\le n-1$.
Hence we can rewrite (\ref {proj}) in the form
\begin{equation}\label{tr_proj}
{\mathbb E}(x)=\frac {1}{n}\sum \limits _{p,q=0}^{n-1}Tr(x_{qp}x)x_{pq},\ x\in B(H).
\end{equation}

Let us consider the family of projections $Q_s$ on the subspaces $F_s\subset H$ spanned by the vectors
$$
(I\otimes S^k)\ket {ss},\ 0\le k\le n-1.
$$

 {\bf Theorem 1.} {\it 
 $$
 \mathbb {E}(Q_j)=\frac {1}{n}I
$$
}

{\bf Proof.} 
$$
x_{p,q}Q_s=\sum \limits_{k=0}^{n-1}\ket{h_k^p}\bra{h_k^q} \sum \limits_{l=0}^{n-1}\ket{s\ s+l}\bra{s\ s+l}=
$$
$$
\frac{1}{n} \sum \limits_{k=0}^{n-1}(\sum \limits_{r=0}^{n-1} e^{\frac{2\pi {\bf i} k r}{n}}\ket{r\ r+p}) (\sum \limits_{m=0}^{n-1} e^{-\frac{2\pi {\bf i} k m}{n}}\bra{m\ m+q}) \sum \limits_{l=0}^{n-1}\ket{s\ s+l}\bra{s\ s+l}=
$$
$$
\frac{1}{n} \sum \limits_{k=0}^{n-1}\sum \limits_{r=0}^{n-1} \sum \limits_{l=0}^{n-1} \sum \limits_{m=0}^{n-1} e^{\frac{2\pi {\bf i} k (r-m)}{n}}\braket{m\ m+q|s\ s+l}\ket{r\ r+p}\bra{s\ s+l}=
$$
$$
\frac{1}{n} \sum \limits_{k=0}^{n-1}\sum \limits_{r=0}^{n-1} e^{\frac{2\pi {\bf i} k (r-s)}{n}}\ket{r\ r+p}\bra{s\ s+q}=
$$
$$
\ket{s\ s+p}\bra{s\ s+q}.
$$
Hence,
$$
Tr(x_{pq}Q_s)=0, \textit{if}\  p\neq q,
$$
and
$$
Tr(x_{pp}Q_s)=1.
$$
Applying (\ref{tr_proj}) we get
$$
\mathbb{E} (Q_s)=\frac{1}{n} \sum \limits _{p=0}^{n-1}x_{pp}=\frac{1}{n} \sum \limits _{p=0}^{n-1}\sum \limits _{k=0}^{n-1}\ket{h_p^k}\bra{h_p^k}=\frac{1}{n}I
$$

$
\Box
$

{\bf Corollary.} {\it The non-commutative operator graph
$$
{\mathcal V}_s=span(\pi (g)Q_s\pi (g^{-1}),\ g\in G_n)
$$
has the anticlique 
$$
P_k=\sum \limits _{j=0}^{n-1}\ket {h_k^j}\bra {h_k^j}.
$$
}

{\bf Proof.}

$$
\braket{h_k^j|s\ s+p}=\frac{1}{\sqrt{n}}\sum \limits _{m=0}^{n-1} e^{\frac{-2\pi {\bf i} k m}{n}}\braket{m\ m+j|s\ s+p}=\frac{1}{\sqrt{n}}e^{\frac{-2\pi {\bf i} k s}{n}} \delta_{jp}.
$$
Hence
$$
P_k Q_s P_k=\sum \limits_{p=0}^{n-1}\sum \limits_{m=0}^{n-1} \sum \limits_{j=0}^{n-1}\ket{h_k^j} \braket{h_k^j|s\ s+p}\braket{s\ s+p|h_k^m}\bra{h_k^m}=
$$
\begin{equation}\label{f1}
\frac{1}{n}\sum \limits_{m=0}^{n-1} \sum \limits_{j=0}^{n-1}\delta _{jm}\ket{h_k^j}\bra{h_k^m}=\frac {1}{n}P_k.
\end{equation}

Taking into account that 
$$
\pi (S)P_k\pi (S^*)=P_{k+1\ mod\ n}\ and\ \pi (M)P_k\pi (M^*)=P_k,
$$
we get
$$
P_k\pi (S^pM^q)Q_s\pi (S^pM^q)^{*}P_k=\pi (S^pM^q)P_{k-p\ mod\ n} Q_s P_{k-p\ mod\ n}\pi (S^pM^q)^{*}=\frac{1}{n}P_k.
$$

$
\Box
$ 

\section {A structure of the non-commutative operator graphs ${\mathcal V}_j$.}

Consider special sums of matrix units (\ref {bas1}) and (\ref {bas2})
$$
y_{ml}=\sum \limits _{k=0}^{n-1}\ket {h_m^k}\bra {h_l^k}.
$$
and their linear combinations
\begin{equation}\label{element}
h_0=\sum \limits _{m=0}^{n-1} y_{m m }
\end{equation}
\begin{equation}\label{elements}
h_p=\sum \limits _{m=0}^{n-1} (y_{m+p \ mod\ n \  m } + y_{m\ m+p \ mod\ n }), 1\le p \le n-1
\end{equation}

{\bf Theorem 2.} {\it All the graphs coincide ${\mathcal V}_j\equiv \mathcal V$. The graph $\mathcal V$ is generated by
the matrices (\ref {element}) and (\ref {elements}).
}

Proof.

It follows from (\ref {bas1}) and (\ref {bas2}) that
\begin{equation}\label{f2}
\ket {s\ s+j}=\frac {1}{\sqrt n}\sum \limits _{m=0}^{n-1}e^{-\frac {2\pi {\bf i} ms}{n}}\ket {h_m^j}.
\end{equation}
Hence
$$
Q_j=\sum \limits _{k=0}^{n-1}\ket {j\ j+k}\bra {j\ j+k}=
\frac {1}{n}\sum \limits _{k=0}^{n-1}\sum \limits _{m=0}^{n-1}\sum \limits _{l=0}^{n-1}e^{\frac {2\pi {\bf i}j(l-m)}{n}}\ket {h_m^k}\bra {h_l^k}=
$$
\begin{equation}\label{Q}
\frac {1}{n}\sum \limits _{m=0}^{n-1}\sum \limits _{l=0}^{n-1}e^{\frac {2\pi {\bf i}j(l-m)}{n}}y_{ml}.
\end{equation}
From (\ref{Q}) we get the following equality
$$
\pi(M^p)\pi(S^q) Q_j\pi (M^q)^*\pi (S^p)^*=\frac{1}{n}\pi(M^p)\pi(S^q) \sum \limits _{m=0}^{n-1} \sum \limits _{l=0}^{n-1}e^{\frac {2\pi {\bf i}j(l-m)}{n}}y_{ml}\pi (M^q)^*\pi (S^p)^*=
$$
$$
 \frac{1}{n}\sum \limits _{m=0}^{n-1} \sum \limits _{l=0}^{n-1}  e^{\frac {2\pi {\bf i}j(l-m)}{n}}e^{\frac {2\pi {\bf i}p(m-l)}{n}}\sum \limits _{k=0}^{n-1}\ket {h_{m+q}^k}\bra {h_{l+q}^k}=
$$
$$ 
 \frac{1}{n} \sum \limits _{m=0}^{n-1} \sum \limits _{l=0}^{n-1}e^{\frac{2\pi {\bf i} (m-l)(p-j)}{n}} y_{m+q\ mod\ n\   l+q\ mod\ n}
$$
It results in
$$
\mathcal{V}_j=span\{z_{qp}^j, 0 \le p,q\le n-1 \},
$$
where
$$
z_{qp}^j=\sum \limits _{m=0}^{n-1} \sum \limits _{l=0}^{n-1}e^{\frac{2\pi {\bf i} (m-l)(p-j)}{n}} y_{m+q \ mod\ n\   l+q \ mod\ n}.
$$ 
Suppose that $j_1 \neq j_0$. Let us take a generator $z_{q_0p_0}^{j_0}$ of $\mathcal{V}_{j_0}$. Then, it coincides with a generator $z_{q_1p_1}^{j_1}$ of $\mathcal{V}_{j_1}$ with parameters
$$
q_1=q_0
$$
and
$$
p_1 =  (p_0 - j_0+ j_1) \ mod \ n
$$
It follows that all the graphs $\mathcal{V}_j$ coincide with 
$$
\mathcal{V}=span\{z_p=\sum \limits _{m=0}^{n-1} \sum \limits _{l=0}^{n-1}e^{\frac{2\pi {\bf i} p(m-l)}{n}} y_{m l }, 0 \le p\le n-1 \},
$$
where $(z_p)$ is a result of the discrete Fourier transform of the system $(h_p)$
$$
z_p=\sum \limits _{m=0}^{n-1} e^{\frac{2\pi {\bf i} pm}{n}}h_m.
$$
By this way,
$$
\mathcal{V}=span\{h_p, 0 \le p\le n-1 \}
$$

$\Box $

\section{Conclusion}

We introduce a number of non-commutative operator graphs generated by reducible unitary representations of the Heisenberg-Weil groups $G_n$. It is shown that all the graphs coincide. The structure of the obtained graph is studied in detail. It is shown that for the graph there exist anticliques. These anticliques are appeared to be projections on the subspaces generated by entangled vectors.
The construction extends one considered in \cite{amosovmokeev2}.

\end{document}